\documentstyle[12pt,aasms4]{article}
\newcommand{\gapr}{\raisebox{-.6ex}{\mbox{
$\stackrel{>}{\mbox{\scriptsize$\sim$}}\:$}}}
\newcommand{\lapr}{\raisebox{-.6ex}{\mbox{
$\stackrel{<}{\mbox{\scriptsize$\sim$}}\:$}}}
\def\einst{{\sl Einstein}}
\def\ros{{\sl ROSAT}}

\def\cxo{{\sl CXO}}

\def\nh{n_{\rm H,20}}
\def\tef{T_{\rm eff}}
\begin{document}
\lefthead{Pavlov et al.}
\righthead{X-ray Spectrum of the Vela Pulsar}
\title{
The X-ray Spectrum of the Vela Pulsar Resolved with {\sl Chandra}}
\author{ 
G.~G.~Pavlov\altaffilmark{1}\altaffiltext{1}{
The Pennsylvania State University, 525 Davey Lab,
University Park, PA 16802, USA; pavlov@astro.psu.edu}, 
V.~E.~Zavlin\altaffilmark{2}\altaffiltext{2}{
Max-Planck-Institut f\"ur Extraterrestrische Physik, D-85740
Garching, Germany; zavlin@xray.mpe.mpg.de}, D.~Sanwal\altaffilmark{1}, 
V.~Burwitz\altaffilmark{2}, 
and G.~P.~Garmire\altaffilmark{1}}
\begin{abstract}

We report the results of the spectral analysis of 
two observations of the Vela pulsar with
the {\sl Chandra} X-ray observatory.  The spectrum of the pulsar
does not show statistically significant spectral lines in the 
observed 0.25--8.0~keV band.
Similar to middle-aged pulsars with detected thermal emission, the spectrum
consists of two distinct components.
The softer component can be modeled 
as a magnetic hydrogen atmosphere spectrum ---
for the pulsar magnetic field $B=3\times 10^{12}$ G
and neutron star mass $M=1.4 M_\odot$ and radius $R^\infty =13$ km, we obtain
$\tef^\infty =0.68\pm 0.03$~MK,
$L_{\rm bol}^\infty = (2.6\pm 0.2)\times 10^{32}$~erg~s$^{-1}$,
$d=210\pm 20$~pc (the effective temperature, bolometric luminosity, and radius
are as measured by a distant observer).
The effective temperature 
is lower than that
predicted by standard neutron star cooling models. 
A standard blackbody fit gives
$T^\infty =1.49\pm 0.04$~MK, 
$L_{\rm bol}^\infty=(1.5\pm 0.4)\times 10^{32}~d_{250}^2$~erg~s$^{-1}$
($d_{250}$ is the distance in units of 250 pc);
the blackbody temperature corresponds to a radius,
$R^\infty =(2.1\pm 0.2)~d_{250}$~km, much smaller than realistic
neutron star radii.
The harder component can be modeled as a power-law spectrum, with parameters
depending on the model adopted for the soft component ---
$\gamma=1.5\pm 0.3$,                                                   
$L_x=(1.5\pm 0.4)\times 10^{31}~d_{250}^2$~erg~s$^{-1}$ and
$\gamma=2.7\pm 0.4$,
$L_x=(4.2\pm 0.6)\times 10^{31}~d_{250}^2$~erg~s$^{-1}$ 
for the hydrogen atmosphere and blackbody soft component, respectively
($\gamma$ is the photon index, $L_x$ is the luminosity in the 0.2--8~keV
band). The extrapolation of the power-law component of the former fit 
towards lower energies
matches the optical flux at $\gamma
\simeq 1.35$--1.45.
\end{abstract}
\section{Introduction}
X-ray observations of rotation-powered pulsars with the \einst, 
{\sl EXOSAT} and, particularly,
\ros\ observatories 
have established (\"Ogelman 1995)
that at least three middle-aged pulsars, PSR B0656+14, B1055--52 and
Geminga, with characteristic ages of
$10^5$--$10^6$~yr, show thermal soft X-ray 
radiation, with (blackbody) temperatures 0.5--1~MK,
interpreted as emitted from the surfaces of cooling
neutron stars (NSs). Investigation of this radiation is 
needed to understand the NS cooling history 
and to study the properties of the NS surface layers.
Further observations with the {\sl ASCA} observatory 
have confirmed 
that, as it had been suggested from the \ros\ data,
the spectra of these pulsars
at higher energies ($\gapr 1.5$--2~keV) are dominated
by nonthermal, power-law components,
with photon indices $\gamma\approx 1.3$--1.6,
presumably generated in the NS magnetospheres
(e.g., Wang et al.~1998). Studying the power-law components
is important to understand 
the mechanisms of the multiwavelength (optical through gamma-rays) pulsar 
radiation.

In addition to these three pulsars (dubbed ``The Three Musketeers'' by 
Becker \& Tr\"umper 1997), the (presumably) much younger Vela pulsar
($\tau=P/[2\dot{P}] = 11$~kyr,
$P=89.3$~ms, $\dot{E}=
6.9\times 10^{36}$~erg~s$^{-1}$, $B\sim 3\times 10^{12}$~G --- see
Taylor, Manchester, \& Lyne 1993 for other properties and references)
has often been mentioned as a possible source of thermal radiation,
based on softness of its spectrum (e.g., \"Ogelman 1995).
\"Ogelman, Finley, \& Zimmermann (1993)
analyzed the spectrum of the Vela pulsar 
detected with 
\ros.
They obtained acceptable fits of the spectrum with
a power law (PL) model 
with $\gamma\simeq 3.3$
and a blackbody (BB) model 
of temperature $T^\infty\simeq 1.7$~MK.
The slope of the spectrum obtained in the PL
fit is much steeper
than typical nonthermal spectra of young and middle-aged pulsars.
The temperature and luminosity inferred from the BB
fit
correspond to an effective radius of 
emitting region about 1.3~km
at a distance $d=250$~pc (Cha, Sembach, \& Danks 1999).
A two-component fit with a BB+PL
model 
yielded a lower temperature, 1.3~MK, a larger radius, 3~km at 250~pc,
and a surprisingly
small photon index, $\gamma\sim 0.1$ (\"Ogelman 1995).
Page, Shibanov, \& Zavlin~(1996) fit the same spectrum with the
NS hydrogen atmosphere models (Pavlov et al.~1995)
and obtained a lower effective temperature and a larger $R/d$:
$T_{\rm eff}^\infty \equiv g_r T_{\rm eff} \simeq 0.8$~MK,
$d\simeq 300$~pc for $R^\infty \equiv g_r^{-1} R =13$~km 
($g_r\equiv [1-2GM/Rc^2]^{1/2}$ is the gravitational redshift parameter).

However, the true spectrum of the observed
soft X-ray pulsar radiation (not to mention the presence of two components)
 has remained elusive because angular resolution
of the X-ray telescopes has been too low to separate 
the pulsar radiation from that of the 
bright pulsar-wind nebula (PWN) of about $2'$ diameter around the pulsar
(Harnden, Grant, \& Seward 1985;
Markwardt \& \"Ogelman 1988).
The subarcsecond resolution of the
 {\sl Chandra} X-ray Observatory (\cxo)
provides the first opportunity to resolve the pulsar spectrum.
The high energy resolution of the \cxo\ grating spectrometers allows one
to address another 
important problem --- if the soft component of the pulsar
radiation indeed originates from the NS surface layers (an atmosphere),
it may show spectral features (lines and photoionization edges) which can
be used to investigate chemical composition, gravity and magnetic field 
of the surface layers. 

In this Letter we present first results on the X-ray spectrum
of the Vela pulsar obtained with \cxo.
We describe the observations 
and the spectral analysis in \S 2 and discuss some
implications of the observed spectrum in \S 3.

\section{Observations and Spectral Analysis}
To investigate the low-energy part of the 
pulsar spectrum with high energy resolution, 
the spectrum dispersed with the Low-Energy Transmission Grating (LETG;
see Brinkman et al.~2000)
was imaged on
the Spectroscopic Array of the High Resolution Camera (HRC-S;
see Murray et al.~1997).
The 25.6~ks observation (ObsID~127) was taken on 28 January 2000.
 For the analysis we used the Level~2 data files 
produced by the standard pipeline processing (v.~R4CU5UPD11.1).
The dispersed LETG spectrum (dispersion 1.148 \AA/mm) was extracted 
from a strip of a $3\farcs6$ (0.18~mm)  
width in the cross-dispersion direction,
which contains about 90\% of source events dispersed
at a given wavelength  
(see {\sl Chandra} Proposers' Observatory Guide, v.3.0; CPOG hereafter).
The zero-order and dispersed radiation from the PWN
contaminates the dispersed pulsar spectrum at short
wavelengths, $\lambda < 5$--6~\AA, and the plate background dominates
at $\lambda > 60$~\AA.
Since at $\lambda = 50$--60~\AA\ there are data only from the
positive dispersed orders, we 
chose the range $\lambda = 6$--50~\AA\ ($E=0.25$--2.0~keV)
for spectral analysis.
The background
was taken from 
boxes
between 
$10''$ and $30''$ (0.49--1.46~mm) from the center of the source spectrum
in the cross-dispersion direction.
To search for spectral lines in the dispersed pulsar spectrum,
we binned the extracted source-plus-background and
background spectra in 
0.02~\AA\ bins (intrinsic resolution
of the instrument in the chosen wavelength range is $\simeq 0.05$~\AA). 
We used groups of 
10--20 sequent bins to estimate
deviation in number of source counts in each bin of a given group 
from the mean value in the group. The maximum deviation,
in both the positive and negative dispersed orders,
 was found to be at a $2.7\sigma$ significance 
level. Therefore, we conclude that the 
LETG  spectrum of the Vela pulsar shows
no statistically significant spectral lines.

 For further analysis, we removed contribution of the higher orders
at longer wavelengths using a ``bootstrap''
method which works as follows 
(see also CPOG).
One assumes that contamination from the higher
orders can be neglected at the shortest wavelengths.
The source counts in a short wavelength bin
are used to deduce the corresponding
high-order contributions of this bin to longer wavelengths
with the aid of relative grating efficiencies of the higher orders.
These contributions are subtracted, and the process is repeated working 
upwards in wavelength.
Using this method, we obtained the first-order 
pulsar spectrum (91\% of the total dispersed data), with 
a source countrate 
$281\pm 6$~ks$^{-1}$ ($312\pm 6$~ks$^{-1}$ after correcting for the
extraction efficiency) in the 0.25--2.0~keV range

To examine 
various
continuum models, we grouped the counts in 162 energy 
bins with at least 30 source counts per bin
and fit the models making use of the first-order effective area\footnote{
version of 31 Oct 2000;
{\tt http://asc.harvard.edu/cal/Links/Letg/User/Hrc\_QE/EA/}}.
We found that a simple PL model 
requires a large photon index $\gamma= 4.1$--4.3 and 
an implausibly high hydrogen
column density $\nh\equiv n_{\rm H}/10^{20}~{\rm cm}^{-2}=9.1$--9.3; moreover,
the best-fit model spectrum exceeds the observed one at $E\gapr 1.2$~keV.
The spectrum fits much better with thermal models (blackbody, NS hydrogen 
atmosphere),
but the observed spectrum somewhat exceeds the model spectra at the highest 
energy channels (1.5--2.0 keV), 
which indicates the presence of a second component with a harder
spectrum. 

To search for the harder component,
we used the archived 37.0~ks observation\footnote{ 
Preliminary results of this observation
were presented 
by Stage et al.~(2000).} (ObsID 131; pipeline processing v.~R4CU5UPD8.2)
taken on 11--12 October 1999
with the Spectroscopic Array of the
 Advanced CCD Imaging Spectrometer (ACIS-S; Garmire et al.~2001)
 in Continuos Clocking (CC) mode which allows timing 
at the expense of one dimension of spatial resolution (along the
CCD chip columns). This observation was carried out with the High Energy 
Transmission
Grating (HETG), which, in principle, gives a high-resolution source spectrum 
at 
higher energies. However, the dispersed spectra,
integrated along the chip columns
in the CC mode, are severely contaminated by the chip background, which 
strongly
complicates their analysis.
Therefore, we use only the zero-order image (on the back-illuminated
ACIS chip S3) which
provides a pulsar spectrum 
with resolution of about 0.1--0.2 keV. The 
archival pipe-line processed data
are not corrected properly for the satellite wobbling (dither) and for the
Scientific Instrument Module motion, which smears the 1-D image.
We applied these corrections making use of the algorithm kindly provided
by Glenn Allen (see Zavlin et al.~2000 for details) and obtained 
one-dimensional (1-D) 
brightness distributions
of the Vela pulsar and its PWN (bottom panel of Fig.~1).
The upper panel of Figure 1 shows a 2-D 
image of this object
obtained with the ACIS-S in the standard Timed Exposure mode 
(ObsID 128)\footnote{Preliminary results of this observation were presented
by Pavlov et al.~(2000). See also 
{\tt http://chandra.harvard.edu/press/00\_releases/press\_060600vela.html}}.
This image is scaled and oriented in such a way that its horizontal
axis is parallel to the 1-D image and its horizontal size is equal to
the length of the 1-D distributions shown in the bottom panel.
The 1-D profiles in different energy bands demonstrate that
the pulsar is much brighter than the PWN background
at $E\lapr 1.5$~keV, and it is discernible up to 8~keV, where the charged 
particle background takes over.

 For the analysis of the pulsar spectrum, 
we extracted counts from a 1-D 2\farcs5-wide aperture.
The nonuniform (in both the total brightness and spectrum) PWN background
was found by interpolating between the
adjacent regions 
on both sides  of the pulsar. 
The estimated source count rate is $288\pm 4$~ks$^{-1}$,
or $320\pm 5$~ks$^{-1}$ after correcting for the 90\% 
fraction of point source counts in the 1-D aperture chosen.
Because of the dither, the pulsar and the PWN were imaged on two nodes
of the chip (node~0 and node~1), which use different amplifiers and
have different gains and responses. Therefore, we 
consider the two nodes as 
separate instruments while fitting
the observed spectra.
Since the ACIS response below 0.5~keV is still poorly known, we 
discarded the corresponding counts.

We find that the ACIS spectrum of the pulsar does not fit with a one-component 
model,
neither standard (BB, PL) nor a NS atmosphere spectrum, but it
fits fairly well with two-component models (e.g., a PL
plus 
a thermal component).

The spectral parameters obtained from fitting the separate HRC-S/LETG and 
ACIS-S
spectra, as well as from the combined fit, are presented in Table~1.
While fitting the LETG spectra, one cannot reliably find the parameters of the
harder component, so we fix them at the values found from the ACIS-S fits.
Similarly, the hydrogen column density $n_{\rm H}$ cannot be determined 
reliably
from the ACIS-S spectra with
low-energy counts discarded, so we fix $n_{\rm H}$ at the value found from the 
LETG fit.
We see from Table~1 that the spectral parameters obtained from the separate
fits are generally consistent with each other; the differences can be
attributed to systematic errors caused by inaccuracies of the HRC/LETG
and ACIS responses.
The combined (LETG plus ACIS) fits (see Fig.~2)
can be considered acceptable,
in view of the current uncertainties of the instrument responses.
Although we cannot completely exclude the possibility that the deviations
from the continuous spectra might be due to some broad, shallow spectral 
features,
the shortness of the exposures and
the current knowledge of the instrument responses do not allow us
to make definitive conclusions based on the data available.
\section{Discussion}
Thanks to the superb angular resolution of the \cxo\ telescope, 
we are able to resolve the pulsar from its PWN and investigate
its spectrum in the 0.25--8.0~keV range. Our analysis proves that
the pulsar soft X-ray emission, at $E\lapr 1.8$~keV, is indeed dominated
by a soft thermal component, so that the Three Musketeers (see Introduction)
are now joined by 
D'Artagnan -- the Vela pulsar.
The parameters of the soft thermal component are drastically different
for the BB fit\footnote{
$T_{\rm bb}^\infty$ and 
$R_{\rm bb}^\infty$ are the temperature and
the radius given by the standard BB
fits; we use the superscript 
$^\infty$
by analogy with the case of atmosphere fits, to emphasize that in both cases
the quantities are given as measured by a distant observer. It should be noted
that $\tef^\infty$ and $R^\infty/d$ very weakly depend on the input parameters
of the hydrogen atmosphere models (Zavlin, Pavlov, \& Tr\"umper 1998). }
($T_{\rm bb}^\infty=1.4$--1.5~MK,
$R_{\rm bb}^\infty=
[1.9$--$2.4] d_{250}$~km --- close to the values obtained by \"Ogelman et 
al.~1993
from the \ros\ spectrum)
 and the hydrogen atmosphere fit
($T_{\rm eff}^\infty =0.65$--0.71~MK, $R^\infty =
[14$--$17] d_{250}$~km).
 For the BB
model, one might assume 
that the observed radiation is emitted
from small hot spots (e.g., polar caps heated by relativistic particles
produced in the pulsar magnetoshere), whereas the $R/d$ ratio obtained
in the H atmosphere fit implies radiation emitted from the entire NS
surface, with a lower temperature.

Rigorously speaking, the surface (atmosphere) of a NS
is not a black body, but a BB
spectrum could mimic the spectrum of a heavy-element (e.g., iron) atmosphere
if it is observed with low spectral resolution 
(Rajagopal, Romani, \& Miller 1997).
However, a lack of significant
spectral lines in the LETG spectrum of the Vela pulsar
hints that there are no heavy elements in the radiating atmosphere
layers (although the exposure was too short to take full advantage
of the high spectral resolution).

Because of the enormous NS gravity, the outer layers of the NS atmosphere
should be comprised of the lightest element present. If, in the absence
of H, the NS has a He atmosphere, we still expect spectral features
in the observed range if the magnetic field is as strong as predicted
by the standard radio pulsar models, $B\sim 3\times 10^{12}$ G.
In such a field the He atmosphere
is not completely ionized even at $T\sim 1$ MK, mainly because of the
large increase of the ionization potentials 
(Ruder et al.~1994): $I\simeq 0.63$~keV
for the one-electron He ion ($\approx 0.5$ keV
with account for the gravitational redshift). The fact that we see neither
a photoionization edge around $E\sim 0.5$ keV nor spectral lines at somewhat
lower energies suggests that there is no He in the atmosphere. On the other 
hand,
if even a small amount of H is present in the surface layers ($\sim 
10^{12}-10^{13}$~g 
of H over the NS surface
 would be enough to hide heavier elements completely), the hydrogen,
with its lower ionization potential, $I \simeq 0.23$ keV, should be strongly
ionizied, and if even some neutral fraction is present, it will not show
spectral features at $E\gapr 0.17$ keV. Thus, the featureless spectrum 
we observe is consistent with the hypothesis of a hydrogen NS atmosphere.
The observed part of the (continuum) spectrum of such an atmosphere
decreases with $E$ slower than the ideal Wien spectrum because
the radiation at higher energies comes from hotter layers
(the H opacity strongly decreases with frequency).
As a result, BB
fits
to the H atmosphere spectra give a temperature higher
than the true effective temperature and an area smaller
than the true emitting area 
(e.g., Pavlov et al.~1995).
If we adopt the H atmosphere hypothesis,
the effective temperature of the Vela pulsar is below the predictions
of the so-called standard models of the NS cooling (e.g., Tsuruta 1998),
if even the pulsar is a factor of 2--3 older than its characteristic age
(Lyne et al.~1996).

An important result of our analysis 
is the detection of the hard spectral component.
Because of the small number of high-energy photons detected,
both the PL and thermal fits of this component are formally acceptable.
However, the thermal fits yield an implausibly small size, $\sim 10$~m,
and a very high temperature, $\sim 10$~MK, of the emitting region,
and extrapolation of the model spectrum to lower and higher energies
is inconsistent with optical and hard-X-ray observations. Therefore,
we favor the nonthermal interpretation of the hard component --- 
a PL ($\gamma=1.2$--1.8
 if the H atmosphere model is adopted for the soft
component) which dominates at $E\gapr 1.8$~keV.
With this interpretation, 
the overall X-ray spectrum of the the Vela pulsar is 
quite similar to those of the Three Musketeers, which show the nonthermal 
component
dominating above 1.5--2.0 keV, with $\gamma \approx 1.3$--1.6.
It is interesting that the ratio of the nonthermal X-ray luminosity
(estimated as if the radiation were isotropic, $L_x=4\pi d^2 F_x$) to 
the spin-down energy loss,
$L_x/\dot{E} =
(1.7$--$2.6)\times 10^{-6}$ in the 0.2--8.0~keV range, 
or 
$(2.1$--$3.4)\times 10^{-6}$ in the
\ros\ range 0.1--2.4 keV, is much smaller than 
$L_x/\dot{E}\sim 10^{-3}$ for majority of X-ray
detected pulsars (Becker \& Tr\"umper 1997).
Thus, luckily for us,
the magnetosphere of the Vela pulsar
is very underluminous in the X-ray range
--- if its $L_x/\dot{E}$ were as high
as for other pulsars, the thermal radiation would be undetectable
in the phase-integrated spectrum.

Since the Vela pulsar has been detected in the optical and gamma-ray
ranges, where its radiation is certainly nonthermal,
 it is illuminating to compare its X-ray nonthermal spectrum with those
in the other ranges. Figure 3 shows the energy flux spectrum of the pulsar
from optical to gamma-rays. We see that extrapolation of the PL
spectrum obtained in the atmosphere+PL
fit matches fairly well
with the optical and hard-X-ray (soft-$\gamma$-ray) time-averaged fluxes.
On the contrary, the slope of the PL
component in the BB+PL
fit is too steep to be consistent with the optical and gamma-ray data,
which is an additional argument in favor of the H atmosphere model.
Moreover, if we adopt the hypothesis that the spectrum has the same slope
in the optical through X-ray range, we can include the optical points
in the fit for the PL
component, 
which constrains the
photon index and the power-law normalization very tightly: $\gamma = 
1.35-1.45$,

In the present paper we analyze only the phase-integrated spectra.
Energy-integrated light curves, with at least three peaks per period,
have been observed with the \cxo\ HRC-I detector
(Pavlov et al.~2000; Helfand, Gotthelf, \& Halpern 2000).
In fact, the phase-resolved spectral analysis, or energy-resolved
timing, can provide much more information about the pulsar. For instance, 
we should expect that the pulsed fraction, pulse shapes (and, perhaps, 
even the number of peaks per period) are quite different at energies
below and above 1.8~keV because of the different emission mechanisms. 
However, because of the low pulsed fraction and complicated light curves,
such an analysis would require an order of magnitude more pulsar photons
than 16,000 photons collected in the described observations.
\acknowledgements
We are grateful to Leisa Townsley and George Chartas for the useful advice 
on the analysis of ACIS data.
Our special thanks are due to Glenn Allen, who provided the algorithm to
correct the event times and coordinates in CC mode.
We thank 
Gottfried Kanbach
and Volker Sch\"onfelder for providing the EGRET and COMPTEL fluxes in a 
digital 
form.
GGP, VEZ and DS are thankful to the Institute for Theoretical
Physics (UCSB), where a part of
this work was done.
This research was supported by NASA grants NAG5-7017
and NAS8-38252, SAO grant GO1-2071X,
NSF grant PHY99-07949, and
DLR (Deutsches Zentrum f\"ur
Luft- und Raumfahrt) grant 50.OO.9501.9.
{} 
\newpage
\voffset=-0.4truein
\begin{table}
\caption{
Parameters of the two-component spectral fits, with 1$\sigma$ uncertainties.}
\begin{tabular}{cccc}
\tableline\tableline
Parameter  & HRC-S/LETG  &  ACIS-S &  combined \\
\tableline
  \multicolumn{4}{c}{Blackbody + Power law }\\
$T^\infty_{\rm bb}$   & $1.47\pm 0.06$  & $1.51\pm 0.03$ & $1.49\pm 0.04$ 
\\ 
$R^\infty_{\rm bb}$
				& $2.2\pm 0.3$ & $2.0\pm 0.2$  & $2.1\pm 0.2$ \\ 
$L_{\rm bol}^\infty$
				& $1.6\pm 0.3$ & $1.4\pm 0.3$  & $1.5\pm 0.4$
\\
$\gamma$
                             & (2.2) & $2.2\pm 0.4$  & $2.7\pm 
0.4$  \\
${\cal N}$
				& (6.0)  & $6.0\pm 1.3$  & $10.7\pm 1.0$
\\
$L_{x, {\rm pl}}$
				& (1.8)    & $2.2\pm 0.3$  & $4.2\pm 0.6$  \\
$n_{\rm H,20}$                    & $1.7\pm 0.3$  & (1.7)  & $2.2\pm 0.3$  \\
$\chi_\nu^2$ [d.o.f.]             & 1.0 [159] & 1.3 [113]   & 1.1 [274]  \\
\tableline
  \multicolumn{4}{c}{H atmosphere + Power law}\\
$T_{\rm eff}^\infty$   & $0.67\pm 0.03$   &  $0.69\pm 0.03$  &  $0.68\pm 
0.03$
\\
$d$            
                        & $200\pm 25$  & $220\pm 20$   & $210\pm 20$ 
\\
$L_{\rm bol}^\infty$
				& $3.9\pm 0.3$ & $3.7\pm 0.3$  & $3.8\pm 0.3$ \\
$\gamma$
                         & (1.4)  & $1.4\pm 0.3$  & $1.5\pm 0.3$  \\
${\cal N}$
				& (2.3)  & $2.3\pm 0.7$  & $2.5\pm 0.6$  \\
$L_{x, {\rm pl}}$
				& (0.5)  & $1.3\pm 0.4$  & $1.5\pm 0.4$  \\
$n_{\rm H,20}$            & $3.0\pm 0.3$  & (3.0)  & $3.3\pm 0.3$  \\
$\chi_\nu^2$ [d.o.f.]   & 0.7 [159]  & 1.2 [113] & 1.0 [274]   \\
\tableline
\end{tabular}
\tablecomments{
The values in parentheses were fixed during fitting.
The temperatures $T_{\rm bb}^\infty$ and
$T_{\rm eff}^\infty$ are in million kelvins (MK). 
The radius $R_{\rm bb}^\infty$ (in km) and the luminosities
$L_{\rm bol}^\infty$ and $L_{x,{\rm pl}}$ (in $10^{32}$ erg s$^{-1}$)
are related to $d=250$ pc. 
The nonthermal
luminosities $L_{x, {\rm pl}}$ are
in the bands 0.2--2, 0.5--8, and 0.2--8~keV
for the LETG, ACIS-S, and combined spectra, respectively. 
$\gamma$ and ${\cal N}$ are the photon index and
 normalization constant of the power-law component:
${\rm d}N/{\rm d}E = {\cal N}
 E^{-\gamma}$, $E$ in keV, ${\cal N}$ in $10^{-4}$~photons/(cm$^2$~s~keV).
The hydrogen atmosphere models were calculated for $B=3\times
10^{12}$ G, $M=1.4 M_\odot$, $R=10$ km ($g_r=0.766$, $R^\infty = 13.05$ km).}
\end{table}
\newpage
\figcaption{
{\sl Bottom}: One-dimensional profiles of the Vela pulsar and its PWN in 
different
energy bands. 
 One pixel is equal to 
$0\farcs492$.
{\sl Top}: $98''\times 70''$
ACIS-S3 image of the Vela pulsar and the PWN,
with the horizontal axis
parallel to the orientation of the 1-D image. 
}
\figcaption{
Combined fit of the
HRC-S/LETG and ACIS-S3 count rate spectra 
with the two-component (magnetic hydrogen
NS atmosphere plus power law) model. 
The dotted and dashed lines
show the contributions of the thermal and nonthermal components.
The best-fit parameters are given in Table 1.
}
\figcaption{
Multiwavelength (1 eV -- 10 GeV) energy spectrum of the Vela pulsar.
In addition to the \cxo\ spectrum from this work,
data shown include results from optical (Nasuti et al.~1997; U,B,V fluxes,
without de-reddening),
OSSE (Strickman et al.~1996), COMPTEL (Sch\"onfelder et al.~2000)
and EGRET (Kanbach et al.~1994) observations.
The solid line shows a two-component
(NS atmosphere plus power-law) fit to the observed \cxo\ 0.25--8 keV spectrum
with spectral parameters $\tef^\infty =0.68$~MK, 
$R^\infty =15.5~d_{250}$~km, $\gamma=1.4$,
${\cal N}=2.3\times 10^{-4}$, $\nh=3.2$. 
The dotted lines correspond to same model spectrum corrected
for the interstellar absorption and extrapolated to lower and higher energies.
The dash-dot lines show the 
 extrapolated optical 
and EUV 
spectra as absorbed by the interstellar matter.
The thin solid lines demonstrate the uncertainty  of the inferred power-law 
component.
A typical error of the thermal component is shown by the vertical errorbar
at the maximum of the spectrum.
}
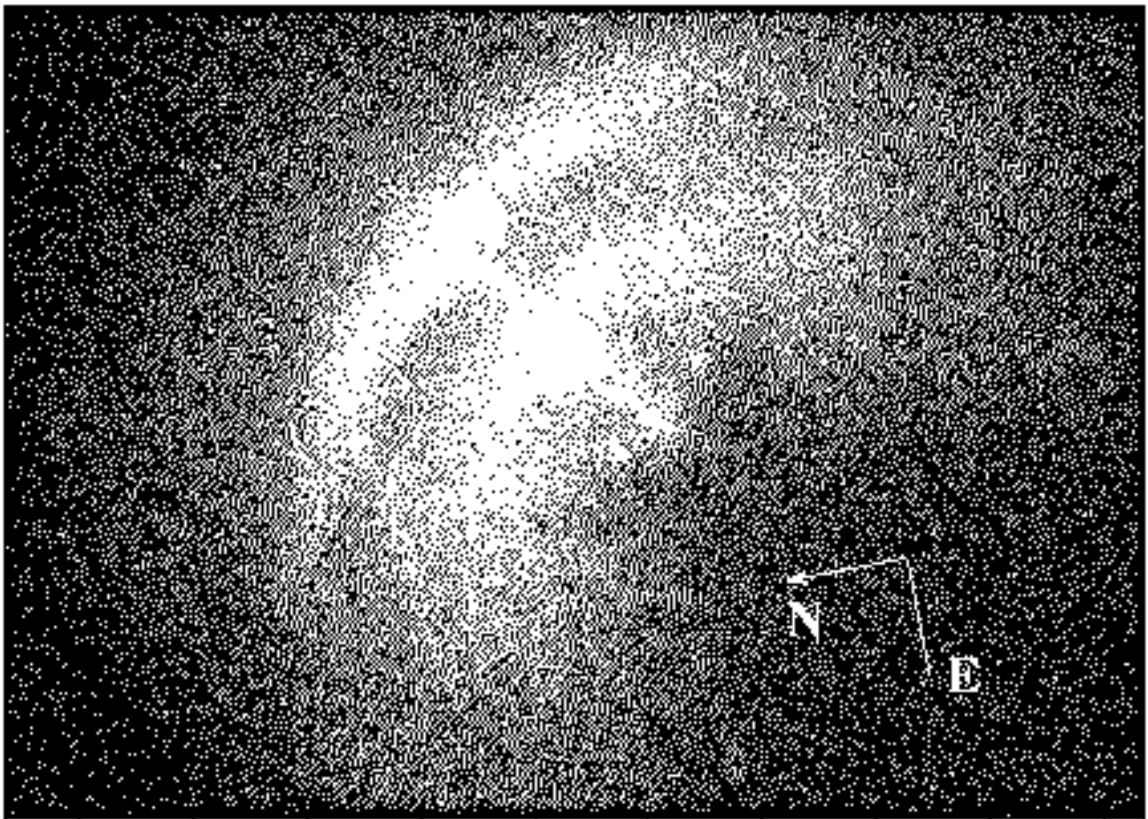
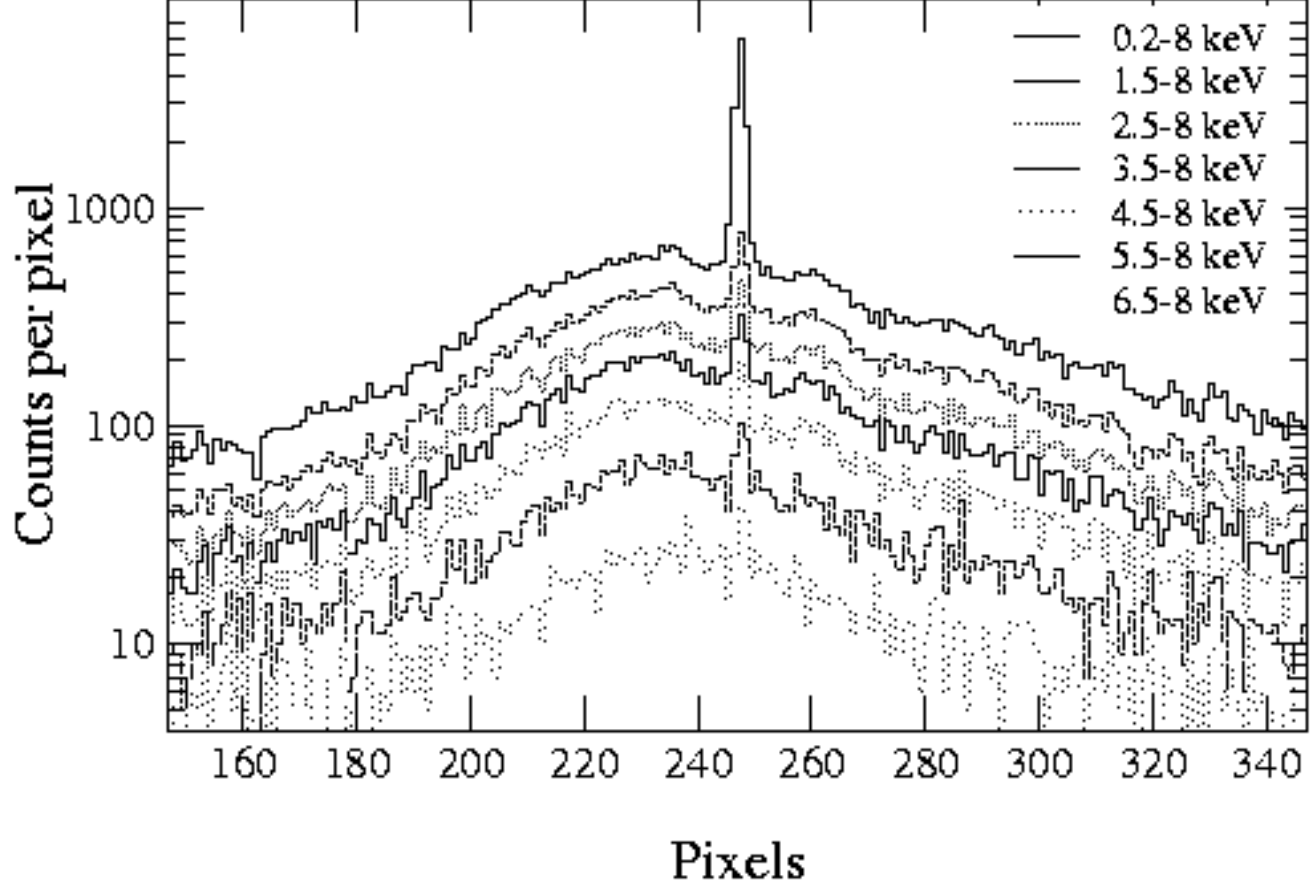
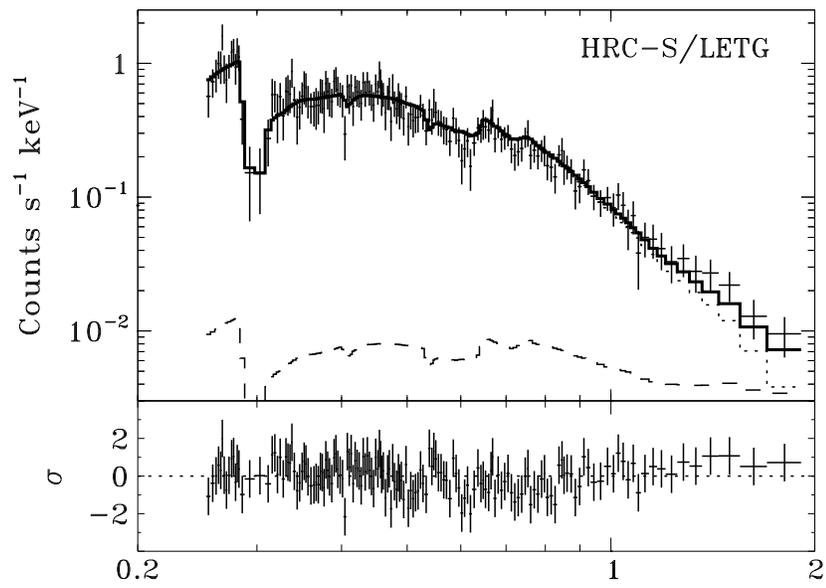
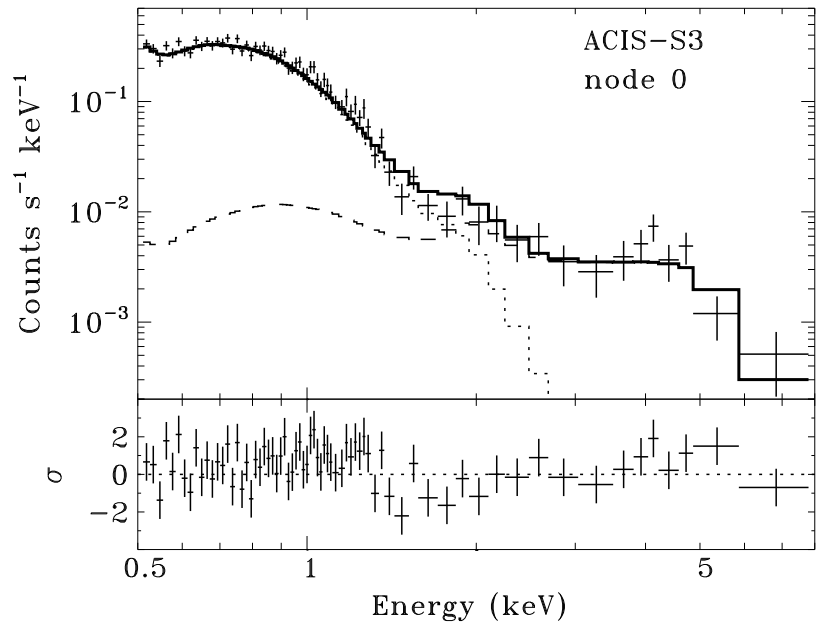
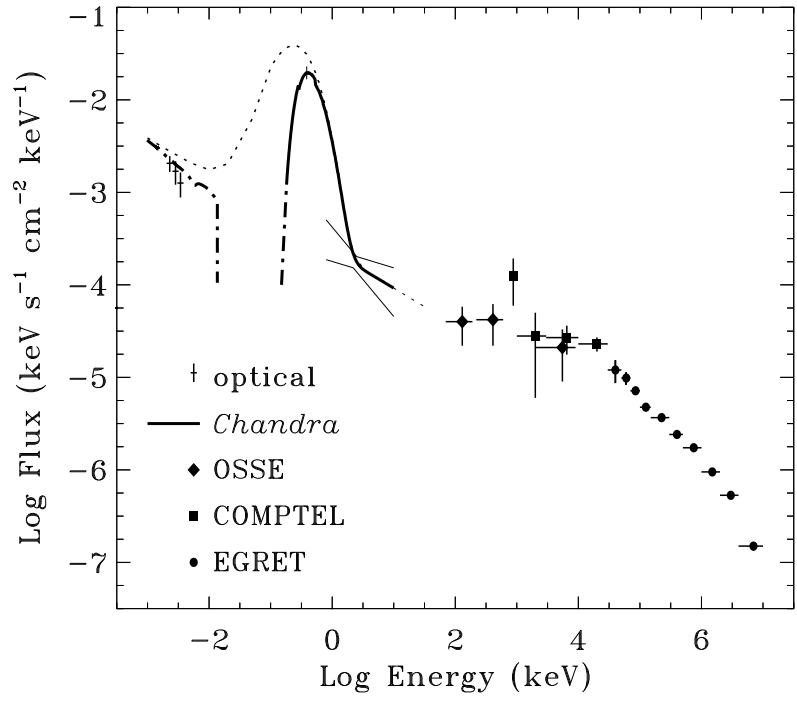
\end{document}